\documentclass[twocolumn,prl,english]{revtex4}
\usepackage{babel}
\usepackage{graphicx}
\usepackage{dcolumn}
\usepackage{bm}

\usepackage[T1]{fontenc}
\usepackage[latin1]{inputenc}
\usepackage{amsmath}
\usepackage{amssymb}
\usepackage{epsfig}
\usepackage{pifont}

\newcommand{\ud}{\mathrm{d}}

\begin{document}
\title{Destructed double-layer and ionic charge separation near the oil-water interface}
\author{Jos Zwanikken, and Ren\'{e} van Roij,}
 \affiliation{Institute for Theoretical Physics, Utrecht University, Leuvenlaan 4, 3584 CE Utrecht, The Netherlands}

\begin{abstract}
We study suspensions of hydrophobic charged colloidal spheres dispersed in a demixed oil-water mixture by means of
a modified Poisson-Boltzmann theory, taking into account image charge effects and partitioning of the monovalent
ions. We find that the ion's aversion for oil can destroy the double layers of the oil-dispersed colloids. This
affects the salt-concentration dependence of the colloidal adsorption to the oil-water interface qualitatively.
The theory also predicts a narrow range of the oil-dielectric constant in which micron-sized water-in-oil droplets
acquire enough charge to crystallize at volume fractions as small as $\sim 10^{-3}$ in the absence of colloids.
These findings explain recent observations [M.E. Leunissen {\em et al.}, Proc. Nat. Ac. Sci {\bf 104}, 2585 (2007)].
\end{abstract} 

\maketitle

Colloidal particles can strongly adsorb to an oil-water or an air-water interface, and hence form a monolayer.
Since a pioneering study by Pieranski \cite{Pieranski} a lot of attention has been devoted to the lateral
structure and the lateral colloidal interactions of such monolayers \cite{Aveyard,Chen}. This two-dimensional
picture is often realistic because the "binding" potential of a single colloid to the interface can be very
strong, typically of the order of $10^3-10^6k_BT$ for a micron-sized colloid due to surface-tension or
image-charge effects \cite{Pieranski,Binks1,Danov}. Here $k_B$ is the Boltzmann constant and $T$ the temperature.
However, in very recent experiments by Leunissen {\em et al.} \cite{Leunissen}, on systems of charged,
micron-sized, hydrophobic poly-methylmethacrylate (PMMA) spheres dispersed in an oil-like mixture of cyclohexyl
bromide and cis-decalin, the focus was not only on the in-plane structure of the adsorbed colloidal monolayer at
the oil-water interface, but also on the out-of-plane "perpendicular" colloidal structure. Striking observations
of this three-dimensional study include (i) a dramatic increase of the lattice spacing of the oil-dispersed
colloidal crystal up to 40 $\mu$m by bringing this dispersion in contact with water, (ii) strong colloidal
adsorption to planar and spherical oil-water interfaces even for non-wetting colloids, (iii) an extremely large
colloid-free zone between the adsorbed monolayer and the bulk crystal in the oil phase, in some cases with a
thickness $d>100\mu$m, and (iv) the existence of micron-sized water-in-oil droplets that are sufficiently charged
to crystallize into a lattice with a spacing of 10-15 microns, without any colloid presence in the system,
provided the dielectric constant of the oil was in the narrow range $4\lesssim\epsilon_o\lesssim 10$
\cite{Leunissen}. In this Letter we develop a theory for a mixture of hydrophobic colloids, cations, and anions in
an oil-water mixture, taking into account screening, image charges, and self energies. We do {\em not} focus on
the lateral structure \cite{Pieranski,Aveyard,Chen} but instead on the structure perpendicular to the interface,
and {\em not} on a single colloid \cite{Danov,Haughey} but on a many-body system. Key new mechanisms that we
identify this way are (i) double-layer destruction, due the preference of ions to reside in the water phase rather
than in diffuse layers surrounding the oil-dispersed colloids, and (ii) ionic charge separation if the cation's
and anion's aversion for oil differs, such that water-in-oil droplets can acquire a net charge. These phenomena
explain essentially all observations of Ref.\cite{Leunissen}, and could be ingredients to further understand and
manipulate Pickering emulsions \cite{Pickering,Binks1,Melle}.

We consider a planar water-oil interface in the plane $x=0$, characterized by a macroscopic surface tension
$\gamma_{wo}$. The interface separates two semi-infinite continuous bulk phases of water ($x<0$) and oil ($x>0$).
This system is a medium for a three-component mixture of hydrophobic colloids (radius $a$, charge $Ze$), cations
(radius $a_+$, charge $e$), and anions (radius $a_-$, charge $-e$). Here $e$ is the elementary charge.  The
strongly hydrophobic character of the colloids is described here phenomenologically through the colloid-water and
colloid-oil surface tensions $\gamma_{cw}=10$mN/m and $\gamma_{co}=1$mN/m, respectively. Note that these tensions
do {\em not} incorporate electrostatic contributions.   Following Pieranski's geometric argument \cite{Pieranski}
a colloidal particle with its center at $x\in(-a,a)$ is therefore subject to the external potential
\begin{equation}\label{fml:Vx}
V_{}(x) = 2\pi a^2 (\gamma_{cw}-\gamma_{co})(1-\frac{x}{a}) -\pi a^2 \gamma_{wo} (1-\frac{x^2}{a^2}), \\
\end{equation}
while a particle completely immersed in water ($x<-a$) or oil ($x>a$) has $V(x)=4\pi a^2
(\gamma_{cw}-\gamma_{co})\simeq 10^6k_BT$ for $a\simeq 1\mu$m and $V(x)=0$, respectively. Note that we shifted the
potential of Ref.\cite{Pieranski} by an arbitrary constant for later convenience. The potential $V(x)$ has a deep
minimum at $x=x^*=a(\gamma_{cw}-\gamma_{co})/\gamma_{wo}\equiv a\cos\theta$ with $\theta$ the wetting angle,
provided $|\gamma_{cw}-\gamma_{co}|< \gamma_{wo}$. Otherwise $V(x)$ is monotonic and we speak of non-wetting.
Below we consider the wetting case $\gamma_{wo}=9.2$mN/m, such that $\cos\theta=0.987$ and $V(x^*)\simeq -10^3
k_BT$, from which strong adsorption of micron-sized colloids at $x\simeq x^*$ is expected. We also consider the
non-wetting case $\gamma_{wo}=9$mN/m such that $\cos\theta=1$, where no strong adsorption is to be expected
because $V(x)$ is monotonic.

The planar water-oil interface also generates an external potential for the ions due to the dielectric
discontinuity, which leads to different electrostatic self energies in the two solvents.  We write the self energy
in medium $i=w,o$ as the Born energy $e^2/(2\epsilon_ia_{\pm})\equiv f_{\pm}(\epsilon_i)$. Denoting the local
dielectric constant by the step function $\epsilon(x)=\epsilon_w$ for $x<0$ and $\epsilon_o$ for $x>0$, this
self-energy effect can be accounted for in terms of (conveniently shifted) ionic external potentials
$V_{\pm}(x)=f_{\pm}(\epsilon(x))-f_{\pm}(\epsilon_w)$, which  vanishes in water and is of the order of
$(1-20)k_BT$ for realistic $\epsilon_o\simeq 4-20$ in the oil phase, i.e. the ions prefer to be in water.

Now that we have specified the external potentials $V(x)$ and $V_{\pm}(x)$ for the colloids and the monovalent
ions, respectively, we employ the framework of density functional theory to calculate the equilibrium density
profiles $\rho(x)$ and $\rho_{\pm}(x)$ \cite{Evans,Zwanikken,Tarazona,Zoetekouw}. The grand-potential functional
$\Omega[\rho,\rho_+,\rho_-]$ is written, per unit lateral area $A$, as
\begin{eqnarray}\label{fml:Omega}
\frac{\Omega}{A}&=&\sum_{\alpha=\pm}\int\!\! \ud x\, \rho_{\alpha}(x)
\Big(k_BT(\ln\frac{\rho_{\alpha}(x)}{\rho_s}-1)+V_{\alpha}(x)\Big)\nonumber\\
&+& \int \ud x\rho(x)\Big(k_BT (\ln\frac{\eta(x)}{\eta_0}-1) +
V(x)\Big)\nonumber\\
&+&k_BT\int \ud x \Big(\rho(x)\Psi(\bar{\eta}(x))+\frac{1}{2}Q(x)\phi(x)\Big),
\end{eqnarray}
where $\eta(x)=4\pi a^3\rho(x)/3$ is the colloidal packing fraction, and where the first and second line are the
ideal-gas grand-potential functionals of the ions and the colloids in their external fields, respectively, and the
third line describes the hard-core and Coulomb interactions. The chemical potentials are represented in terms of
an ion concentration $\rho_s$ and a reference colloid packing fraction $\eta_0$ to be discussed below. The
colloid-colloid hard-core interactions are taken into account by the Carnahan-Starling excess free energy per
particle $\Psi(\bar{\eta})=(4\bar{\eta}-3\bar{\eta}^2)/(1-\bar{\eta})^2$ with the weighted packing fraction
$\bar{\eta}(x)=\int_{x-2a}^{x+2a} dx' w(x-x')\eta(x')$ with the (low-density) weight function
$w(x)=\frac{3}{32}(4a^2-x^2)/a^3$ \cite{Tarazona}. Such a nonlocal treatment of the hard-core interactions is
necessary to describe the extremely localized adsorbed colloidal monolayer in the case of wetting realistically.
The electrostatic interactions between all species are described in Eq.(\ref{fml:Omega}) at a mean-field level in
terms of the total local charge number density $Q(x)=Z\rho(x)+\rho_+(x)-\rho_-(x)$ and the yet unknown
electrostatic potential $k_BT\phi(x)/e$ that must satisfy the Poisson equation and boundary conditions
\begin{eqnarray}\label{fml:PB}
\epsilon(x) \phi''(x) &=& -4\pi\beta e^2 Q(x), \hspace{3mm}\hspace{3mm} (x\neq 0);\\
\lim_{x\uparrow0} \epsilon_w \phi'(x) &=& \lim_{x\downarrow0} \epsilon_o \phi'(x) \hspace{3mm};\hspace{3mm} \lim_{x\rightarrow\pm\infty}
\phi'(x)=0.\nonumber
\end{eqnarray}
Here a prime denotes a derivative with respect to $x$, and $\beta=1/(k_BT)$. Note that the second line of
(\ref{fml:PB}) describes the image-charge effects (first term) and global charge neutrality (second term).
Minimizing the functional leads to the Euler-Lagrange equation
$\delta\Omega/\delta\rho(x)=\delta\Omega/\delta\rho_{\pm}(x)=0$, which can be rewritten as
\begin{eqnarray}\label{fml:Boltzmann}
\eta(x)&=&\eta_0 \exp\big(-\beta V(x)-Z\phi(x)- \beta\mu(x)\big);  \nonumber \\
\rho_\pm(x)&=&\rho_s \exp\big(-\beta V_{\pm}(x) \mp \phi(x)\big),
\end{eqnarray}
with $\beta\mu(x)=\Psi(\bar{\eta}(x))+\int_{x-2a}^{x+2a} dx'w(x-x')\eta(x')\Psi'(\bar{\eta}(x'))$ where
$\Psi'(\eta)=d\Psi(\eta)/d\eta$. Note that the strongly hydrophobic character of the colloids leads to a vanishing
colloid-density at $x\rightarrow-\infty$, such that $\phi(-\infty)=0$ and $\rho_{\pm}(-\infty)=\rho_s$ for electro
neutrality reasons, i.e. the bulk water phase acts as the salt reservoir with a total ion concentration $2\rho_s$.
The corresponding Debye screening length in the water phase is $\kappa_w^{-1}=(8\pi\beta e^2
\rho_s/\epsilon_w)^{-1/2}$. In the bulk oil-suspension, $x\rightarrow\infty$, the average colloid packing fraction
$\eta(\infty)\equiv\eta_b$ can be imposed by tuning $\eta_0$ appropriately. Eqs.(\ref{fml:PB}) and
(\ref{fml:Boltzmann}) also imply, for $x>0$, that $\phi''(x)=\kappa_o^2\sinh(\phi(x)-\phi_c)-4\pi\beta e^2 Z
\rho(x)/\epsilon_o$, with the Debye screening length $\kappa_o^{-1}$ in the oil phase defined by
$\kappa_o^2=\kappa_w^2(\epsilon_w/\epsilon_o)\exp[-\beta(V_+(\infty)+V_-(\infty))/2]$ and
$\phi_c=\beta(V_-(\infty)-V_+(\infty))/2$.

Eqs.(\ref{fml:PB}) and (\ref{fml:Boltzmann}) form a closed set for the four unknown profiles $\rho(x)$,
$\rho_{\pm}(x)$, and $\phi(x)$, and can be solved numerically on an $x$-grid by standard iterative methods on
desktop PC's. Typically we use several hundred non-equidistant grid points, with a relatively small grid spacing
close to $x=0$ and close to the monolayer position $x=x^*$ in the wetting case.

\begin{figure}[!ht]
\centering
\includegraphics[width = 7.3cm, angle = 270]{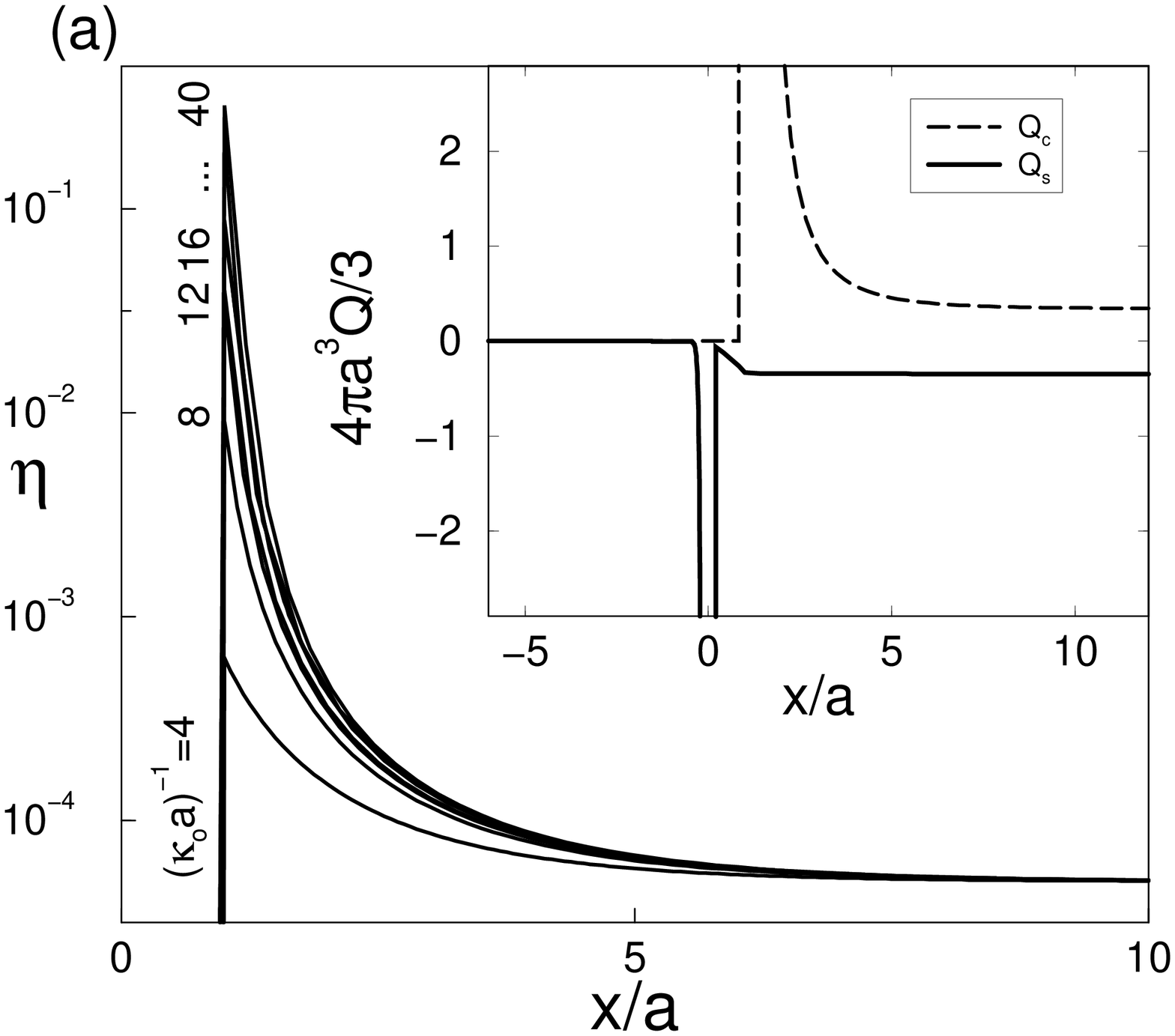}
\includegraphics[width = 7.3cm, angle = 270]{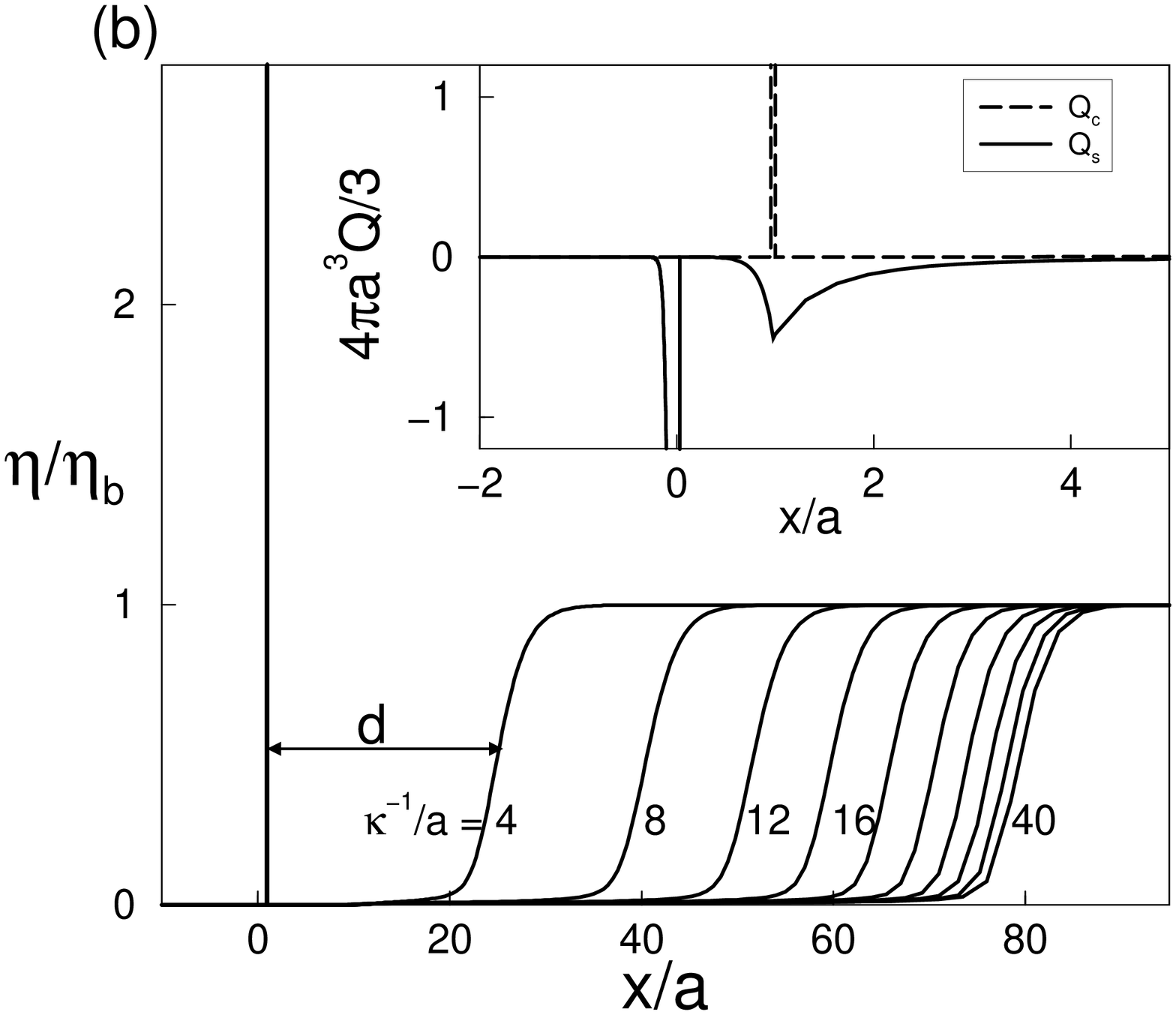}
\caption{The packing fraction profile $\eta(x)$ of strongly hydrophobic, oil-dispersed colloidal spheres (radius
$a=1\mu$m, charge $Z=450$) in the vicinity of a planar interface at $x=0$ between water ($x<0$, dielectric
constant $\epsilon_w=80$) and oil $(x>0, \epsilon_o=5.2)$, for a colloidal bulk packing fraction
$\eta_b=\eta(\infty)=5\times 10^{-5}$, in (a) for nonwetting colloids ($\cos\theta=1$) for oil screening lengths
$\kappa_o^{-1}/a=4-40$ from bottom to top, and in (b) for weakly wetting colloids ($\cos\theta=0.987$) for
$\kappa_o^{-1}/a=4-40$ from left to right.  The insets show, for $\kappa_0^{-1}=40a$, the salt and colloidal
charge distributions, revealing the complete (a) and partial (b) destruction of the double layer around the
adsorbed colloidal layer.} \label{fig:figure1a}
\end{figure}

We fix as many parameters as possible in accordance with those in the experiments described in
Ref.\cite{Leunissen}, such that $Z=450$, $a=1\mu$m, $\eta_b=5\times 10^{-5}$, $\epsilon_w=80$, $\epsilon_o=5.2$.
For simplicity we take equal ionic sizes $a_+=a_-=0.3$ nm such that $V_{\pm}(x)=17k_BT$ for $x>0$ for both ionic
species. This leads to $\kappa_o^{-1}/a=1-40$ if one varies $\rho_s$ from 0.07 to 26 mM. In Fig. \ref{fig:figure1a}
we show, for these parameter choices, the resulting colloidal packing fractions $\eta(x)$  for several screening
lengths $\kappa_o^{-1}$, in (a) for the nonwetting case $\cos\theta=1$  and in (b) for the partial wetting case
$\cos\theta=0.987$.  In all cases the colloids are so hydrophobic that $\eta(x)$ is vanishingly small for $x<0$.
The qualitative difference between the two sets of curves in (a) and (b) is striking, given the small difference
in contact angle. In Fig.1(a) we observe $\eta(x)$ to increase by up to several orders of magnitude near the
interface at $a<x\lesssim 5a$, whereas in Fig.1(b) we observe a densely packed monolayer of colloids at $x=x^*$
separated from the bulk by a colloid-free zone of a thickness $d$ of the order of several $\kappa_o^{-1}$.
Although the increase of $\eta(x)$ up to $10^3\eta_b$ close to the interface in Fig.1(a) is significant, it is
much weaker than to be expected on the basis of the attractive image charge potential
$W(x)=ZZ'e^2/(4\epsilon_ox)\simeq -500 ak_BT/x$ that a single colloid (without cations and anions) would
experience at $x>a$ for the present parameters, with $Z'=-Z(\epsilon_w-\epsilon_o)/(\epsilon_w+\epsilon_o)=-395$
the image charge \cite{Danov}. At first sight one would attribute this relatively modest colloidal adsorption to
the mutual repulsion between adsorbed colloids, which in the present many-body description is included through the
mean-field Coulombic term in Eq.(\ref{fml:Omega}). However, in that case one would expect a stronger
(longer-ranged) colloid-colloid repulsion and hence a weaker colloidal adsorption for increasing $\kappa_o^{-1}$,
whereas Fig.1(a) shows the opposite trend. This implies that another mechanism is at work here. The mechanism that
we identify involves the salt and colloidal charge distributions $Q_s(x)=\rho_+(x)-\rho_-(x)$ and
$Q_c(x)=Z\rho(x)$, respectively, shown in the insets of Fig.1 for $\kappa_o^{-1}/a=40$. These insets show a charge
separation due to the accumulation of a net negative ionic charge in water, in a very thin layer
$-\kappa_w^{-1}<x<0$, and a net positive colloidal charge for $a<x<5a$ in (a) and at $x=x^*$ in (b). In other
words, the usual double-layer structure, in which the compensating ionic charge resides in the vicinity of several
Debye lengths of the colloidal charged surfaces, is destroyed completely (a) or partially (b) for hydrophobic
colloids in the vicinity of the oil-water interface: a fraction of the compensating ionic charge prefers to reside
in the water phase "further away" rather than close to the colloidal surface. Since the length scale for local
charge separation is set by the Debye length, this fraction, and hence the net charge in the thin layer
$-\kappa_w^{-1}<x<0$ in the water, will increase with increasing $\kappa_o^{-1}$. This accumulated charge attracts
the oppositely charged colloids at $x>a$ in Fig.1(a), and hence the adsorption of colloids increases with
$\kappa_o^{-1}$. In a an oversimplified dynamic sense one could envisage a colloid with a spherical double layer
approaching the water-oil interface from the oil side, then being stripped from (part of) its ionic cloud if it
gets to close, such that it cannot diffuse back to the bulk as a neutral entity and hence adsorbs. Double-layer
destruction also takes place in the wetting case of Fig.1(b), where the deep well of $V(x)$ at $x=x^*$ drives the
formation of a close-packed monolayer at $x=x^*$. Due to colloid-colloid repulsions this monolayer strongly repels
the colloids in bulk, giving rise to the colloid-depleted zone of a thickness $d$ that increases with
$\kappa_o^{-1}$, as expected on the basis of this argument, in agreement with Fig.1(b). However, whereas $d\simeq
4\kappa_o^{-1}$ for $\kappa_o^{-1}\lesssim 10a$, $d$ is levelling off for larger $\kappa_o^{-1}$ to a maximum
value of about $d\simeq 80 a$ at $\kappa^{-1}_o=50a$, beyond which $d$ is decreasing with $\kappa_o^{-1}$ (not
shown). This peculiar behavior of $d$ at large $\kappa_o^{-1}$ is entirely due to the double-layer destruction,
which causes (an increasing fraction of) the compensating charge of the monolayer at $x=x^*$ to reside in the
water phase at $-\kappa_w^{-1}<x<0$, such that the geometry tends to be that of two oppositely charged plates
which have a strong field inside but a weaker field outside, i.e. the repulsion between the monolayer and the bulk
colloids is reduced and hence $d$ shrinks. This non-monotonic behavior of $d$ with $\kappa_o^{-1}$ is a mere
prediction of the present theory, but the observations of $d$ decreasing from about 100 to 20 micron upon adding
salt in the water phase \cite{Leunissen} (and hence varying $\kappa_o^{-1}$ over the parameter regime of Fig.1(b))
is in agreement with the theoretical values.

One problem in the comparison between the results of Fig.1 and the experiments of Ref.\cite{Leunissen} is that the
colloid-depleted zone was observed for non-wetting colloids, whereas Fig.1(b) holds for the case of wetting
colloids, $\cos\theta=0.987$. One could on the one hand of course argue that the wetting angle of the experimental
system is smaller than the resolution of its measurement, such that the experimental system would actually be
(extremely weakly) wetting. But on the other hand we found in extended calculations that profiles such as those of
Fig.1(b) can also be obtained for non-wetting colloids provided we take into account the combined effects of (i)
charge regulation, such that an additional profile $Z(x)$ is calculated using ionic association-dissociation
equilibrium as an additional condition, and (ii) a nonlocal analogue of the local relation $Q_c(x)=Z\rho(x)$, such
that a colloid at $x$ contributes to the charge density in the whole interval $[x-a,x+a]$. These extension will be
published elsewhere.

In Ref.\cite{Leunissen} it was also reported that water-in-oil droplets could be stable {\em without} any colloids
or any other additives in the system. This observation goes against the common believe that emulsions require
"emulsifiers" in order to be stable \cite{Binks1998}. The stabilization mechanism proposed in Ref.\cite{Leunissen}
is based on the asymmetry between the cations and anions as regards their self-energy in oil and water, such that
the droplets spontaneously acquire a net charge. Moreover, it was observed that a system of water-in-oil droplets
could actually even crystallize, but only if $4\lesssim\epsilon_o\lesssim10$. The present theory as formulated in
Eqs.(\ref{fml:PB})-(\ref{fml:Boltzmann}) can be employed to underpin and further understand these recent
surprising observations. The absence of any colloids is modelled by setting $\rho(x)\equiv 0$ throughout, and the
alleged asymmetry of the cations and anions is taken into account by setting their size ratio
$a_+/a_-\equiv\xi<1$, such that $V_-(x)=\xi V_+(x)$, i.e. the cations' aversion for oil is larger than the
anions'. For $\xi=0.5$ and $\epsilon_w=80$, we solved Eqs.(\ref{fml:PB})-(\ref{fml:Boltzmann}) on an $x$-grid to
obtain the equilibrium profiles $\rho_{\pm}(x)$, for $\rho_s[M]=10^{-5},10^{-1}$, for a range of $\epsilon_o$ and
$a_+$. The net (positive) charge in the water phase (per unit area, in units of $e$) follows as
$\sigma=\int_{-\infty}^0 dx (\rho_+(x)-\rho_-(x))$. If we now presume that a spherical water-in-oil droplet of
radius $R$ has the same surface charge density, it would have a total charge number $Z=4\pi R^2\sigma$, and two of
these droplets would repel each other by a potential of the form $v(r)= (Ze)^2\exp(\kappa_o(2R-r))/[\epsilon_o
(1+\kappa_oR)^2 r]$ \cite{Zoetekouw}. Inspired by the experiments of Ref.\cite{Leunissen} we set $R=1.5\mu$m, and
consider $\beta v(r)$ for the inter-droplet spacing $r=10R$ as a function of $\epsilon_o$ and $a_+$ as shown in
the contour plot of Fig.2. It is clearly observed that $v(10R)$ only exceeds $1k_BT$ in a very narrow band of
values for $\epsilon_o$. Given that one expects crystallization of a dilute system of charged spheres as soon as
$v(r)$ at typical droplet spacings exceeds ${\cal O}(k_BT)$, the narrow $\epsilon_o$-bands of Fig.2 correlate with
the observations in Ref.\cite{Leunissen}. The peak in $Z$ at $\epsilon_o\simeq 30\gg 10$, shown in the inset of
Fig.2, shows that $v(10R)$ is determined by an intricate balance of amplitude and range; the fact that
$Z\lambda_B/R\lesssim 5$ for $\epsilon_o<10$, with $\lambda_B=\beta e^2/\epsilon_o$ the Bjerrum length of the oil,
indicates that charge renormalization is of {\em no} concern in the regime where $\beta v(10R)\simeq 1$
\cite{Zoetekouw}.

In conclusion, we have presented a theory for charged colloids, anions, and cations near a water-oil interface,
taking into account colloidal wetting properties, ionic self-energies, electrostatic image effects and screening.
We focussed on the structure perpendicular to the interface, and identified the destruction of colloidal double
layers and the separation of ionic charge as important mechanism taking place near this interface. With these
non-lateral mechanisms we can describe several recent observations of hydrophobic charged colloids, most notably
the large colloid-depleted zone in between the adsorbed monolayer and the bulk oil-dispersed colloidal phase, the
spontaneous self-charging of water-in-oil droplets, and their crystallization in a narrow band of $\epsilon_o$
\cite{Leunissen}. An extension of the theory to describe the very recent observations of stable Pickering
emulsions \cite{Sacanna} is in progress.

It is a pleasure to thank M.E. Leunissen and A. van Blaaderen for sharing their experimental results with us before
publication, for useful discussions, and for carefully reading the manuscript.

\begin{figure}[!ht]
\centering
\includegraphics[width = 7.1cm, angle = 270]{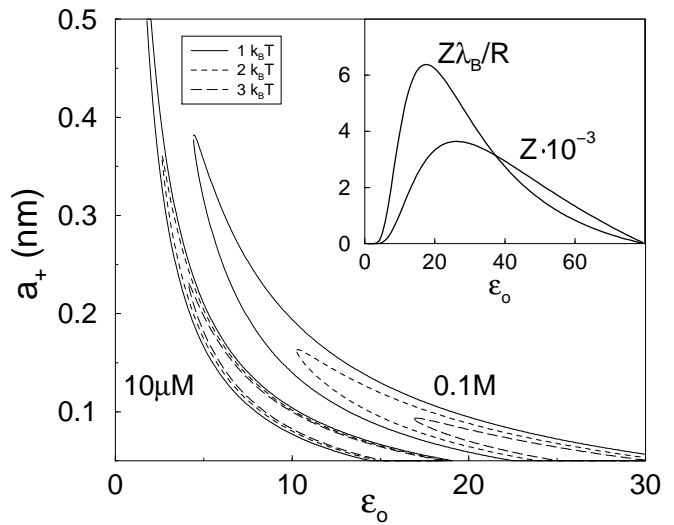}
\caption{Contours of the pair potential $v(r)$ between water-in-oil droplets of radius $R=1.5\mu$m at separation
$r=10R$, as a function of cation radius $a_+$ and dielectric constant of the oil $\epsilon_o$, for two water salt
concentrations $\rho_s=10\mu$M and 0.1M. The anion radius is $a_-=2a_+$ here, and no colloidal particles are
present. Within the similar narrow region of $\epsilon_o$ we find that $v(10R)-v(9R) = \mathcal{O}(k_BT)$.The
inset shows, for $a_+=0.3$nm and $\rho_s=10^{-5}$M, the charge $Z$ and combination $Z\lambda_B/R$ of the droplet
(see text).} \label{fig:figure2}
\end{figure}

\end{document}